\documentclass[conference]{IEEEtran}
\IEEEoverridecommandlockouts
\usepackage{cite}
\usepackage{amsmath,amssymb,amsfonts}
\usepackage{algorithmic}
\usepackage{graphicx}
\usepackage{textcomp}
\usepackage{xcolor}
\newcommand{\todo}[1]{}
\renewcommand{\todo}[1]{{\color{red} TODO: {#1}}}
\newcommand\Tstrut{\rule{0pt}{2ex}}       
\newcommand\Bstrut{\rule[-0.9ex]{0pt}{0pt}} 
\newcommand{\TBstrut}{\Tstrut\Bstrut} 
\def\BibTeX{{\rm B\kern-.05em{\sc i\kern-.025em b}\kern-.08em
    T\kern-.1667em\lower.7ex\hbox{E}\kern-.125emX}}
\begin{document}

\title{Runtime Monitoring of Human-centric Requirements in Machine Learning Components: A Model-driven Engineering Approach   }

\author{\IEEEauthorblockN{Hira Naveed}
\IEEEauthorblockA{\textit{HumaniSE Lab, Department of Software Systems and Cybersecurity, Faculty of IT, Monash University} \\
Clayton, Australia \\
hira.naveed@monash.edu}
}

\maketitle

\begin{abstract}
As machine learning (ML) components become increasingly integrated into software systems, the emphasis on the ethical or responsible aspects of their use has grown significantly. This includes building ML-based systems that adhere to human-centric requirements, such as fairness, privacy, explainability, well-being, transparency and human values. Meeting these human-centric requirements is not only essential for maintaining public trust but also a key factor determining the success of ML-based systems. However, as these requirements are dynamic in nature and continually evolve, pre-deployment monitoring of these models often proves insufficient to establish and sustain trust in ML components. Runtime monitoring approaches for ML are potentially valuable solutions to this problem. Existing state-of-the-art techniques often fall short as they seldom consider more than one human-centric requirement, typically focusing on fairness, safety, and trust. The technical expertise and effort required to set up a monitoring system are also challenging. In my PhD research, I propose a novel approach for the runtime monitoring of multiple human-centric requirements. This approach leverages model-driven engineering to more comprehensively monitor ML components. This doctoral symposium paper outlines the motivation for my PhD work, a potential solution, progress so far and future plans.
\end{abstract}


\section{Introduction}
Software systems with machine learning (ML) components are becoming increasingly prevalent~\cite{nigenda2022amazon,schroder2022monitoring}. With applications in domains such as healthcare, agriculture, finance and e-commerce, several decisions in these systems and their users are being controlled by ML~\cite{pessach2022review}. However, during the development of ML components, critical human-centric requirements, such as fairness, privacy and well-being, are often given low priority or ignored~\cite{heyn2021requirement,mougouei2018operationalizing}. This leads to multiple consequences ranging from privacy violation to bias against users with special needs~\cite{crawford2021hidden,solanki2023operationalising}. Some examples of recent problems found in ML-based software include fairness violations (gender bias) in Amazon's recruitment tool~\cite{dastin2022amazon}, respect and dignity violations in Microsoft's Tay chatbot~\cite{brandtzaeg2018chatbots} and reliability violations due to incorrect diagnosis in IBM Watson's cancer diagnosis system~\cite{strickland2019ibm}.

Addressing the lack of human-centric requirements coverage requires a twofold approach. One, considering human-centric requirements of primary importance in pre-deployment stages like ML model design, creation, training and testing. Two, continuously monitoring the ML components after deployment to ensure they continue to meet human-centric requirements. In this PhD, I focus on the latter.

\begin{figure}
    \centering
    \includegraphics[width=0.48\textwidth]{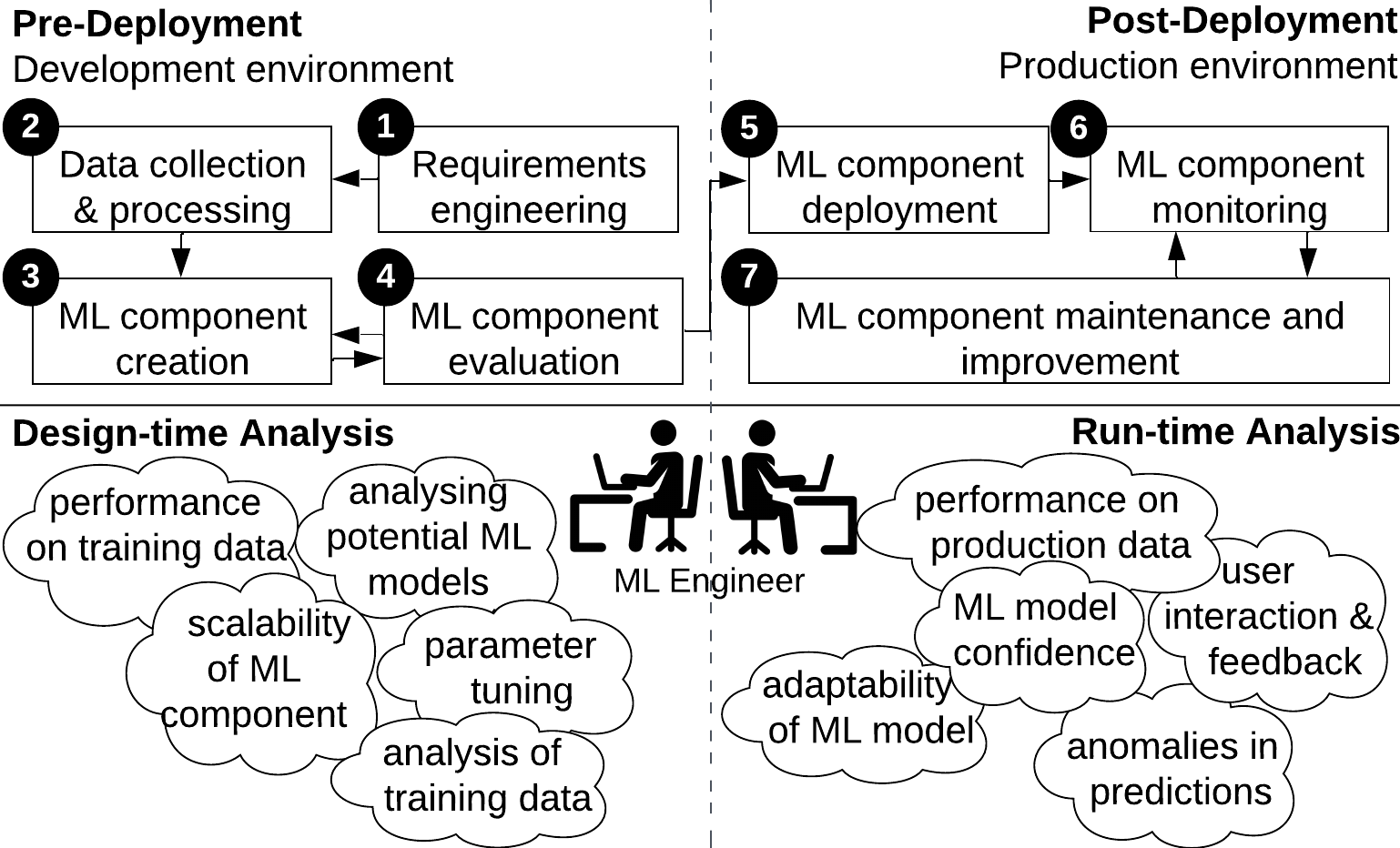}
    \caption{Differences between Design Time and Runtime Analysis}
    \label{fig:analysis}
    \vspace*{-1em}
\end{figure}

Monitoring involves continuously observing the behaviour of an ML component at runtime by capturing and analysing relevant data to identify unwanted behaviour \cite{schroder2022monitoring}. It differs significantly from the analysis done at design time, as shown in Fig.~\ref{fig:analysis} because several properties of the ML component can only be analysed at runtime. For example, the performance and prediction confidence on production data, anomalies in predictions, differences between training and production data and user interaction patterns and feedback. 

~\label{sec:introduction}

\section{The Problem}
The need for runtime monitoring of ML software components arises because conducting verification and validation before deployment is insufficient to guarantee the correct behaviour of ML components \cite{zhu2022ai,guerin2022unifying}. One reason for this is the limitations of the development environment, as shown in Figure 1. For example, the absence of real-world data, changing end-user requirements, changing real-world contexts of use, additional data after deployment, etc. Another reason is the short time to market when developing the ML component. In addition, the autonomous and adaptable nature of ML makes it impossible to accurately and completely specify all the goals in detail \cite{zhu2022ai,guerin2022unifying}. ML components are brittle and may produce incorrect outputs in unexpected production environments \cite{torfahruntime,rabanser2019failing,cummaudo2020threshy, ghosh2022faircanary}. Even subtle differences between training and production data or changes in how variables relate can significantly compromise performance \cite{rabanser2019failing}. Thus, ML components in production must be continuously monitored to ensure desired performance and reliability \cite{nigenda2022amazon,heyn2021requirement}.

Current research and industrial works include several solutions for monitoring deployed ML components~\cite{kourouklidis2020towards,liu2002data,zhou2019framework,guerin2022unifying,lewis2022augur}. However, most works focus on monitoring for performance degradation \cite{kourouklidis2020towards,zhou2019framework} and changes in data distribution \cite{lewis2022augur,liu2002data,eck2022monitoring,aslansefat2020safeml}. Studies that support runtime monitoring of ML components for human-centric requirements other than fairness \cite{nigenda2022amazon,henzinger2023runtime,albarghouthi2019fairness}, trust \cite{langford2021modalas,byun2020manifold,roy2022runtime} and safety \cite{aslansefat2020safeml,guerin2022unifying,torfahruntime} are rare. I motivate my PhD research program by highlighting this gap in the literature and providing an example to explain the importance of this problem and why it needs to be solved. 


Consider the fictional example of an autonomous drone delivery system a company uses to deliver packages to customers' homes. The drone delivery system uses ML components for route optimisation, autonomous navigation, delivery destination recognition and package recognition and handling. This system can violate many human-centric requirements during operation, for example:
\begin{enumerate}
\item  \textbf{Fairness and Inclusion:} The drone delivery system discriminates in delivery prioritisation against certain neighbourhoods or demographic groups, leading to unequal access to services. For example, due to closely packed homes in economically disadvantaged areas, the drone delivery system fails to provide the service.
\item  \textbf{Privacy:} The drone delivery system captures and stores sensitive data. For example, images or videos of people's homes or private property.
\item  \textbf{Safety:} The drone delivery system operates in a manner that compromises safety. For example, the drone flies at high speeds in crowded areas and causes anxiety.
\item  \textbf{Self-direction:} The drone delivery system restricts customers from customising delivery options. For example, the system forces customers into predefined schedules.
\item  \textbf{Pleasure:} The drone delivery system prioritises efficiency at the expense of customer satisfaction. For example, the system tries to minimise delivery time at all costs, leading to rushed operations.
\item \textbf{Honesty: } The drone delivery system prioritises making an income via nontransparent and dishonest pricing, billing and subscription practices. 
\end{enumerate}

These violations can occur because the development team failed to consider many human-centric requirements during development, or because some of these violations emerged later. In either scenario, without runtime monitoring, the system heavily relies on user complaints as the primary means of discovering these issues. This delayed approach reduces users' trust in the system and discourages them from using it in the future \cite{obie2022violation}. Further, it may also cause anxiety, frustration and disappointment since the system does not meet the needs of its target users \cite{grundy2020humanise}. 

~\label{sec:problem}

\section{Expected Key Contributions}
I plan to make the following contributions through my research: 
\begin{itemize}
    \item Identification of key human-centric requirements for ML components that require runtime monitoring.
    
    \item Define a set of domain-specific modeling languages (DSMLs) at different levels of abstraction to model human-centric requirements in ML components, technical requirements of ML components, system architecture, ML component design and system context.
    
    \item Design and implementation of a model-driven approach for generating runtime monitors for human-centric requirements in ML components.

    \item Techniques to monitor changing requirements at runtime.
    
    \item Automated classification mechanism for human-centric requirements violations.
    
    \item Adaptation mechanisms to fix human-centric requirements violations where possible.
    
    \item Tool support for the proposed approach.

    \item Evaluations of my work on real-world case studies with real-world developers.
\end{itemize}
~\label{sec:Contributions}

\section{Related Work}
Generally, runtime monitoring of ML has been studied with a focus on technical aspects like performance drops \cite{eck2022monitoring,buerkle2021fault}, data distribution changes \cite{kourouklidis2020towards,liu2002data,eck2022monitoring,lewis2022augur,bhattacharjee2019stratum,guerin2022evaluation,klaise2020monitoring,kourouklidis2021model,vertexwebsite,azurewebsite} and ML model changes \cite{kourouklidis2020towards,zhou2019framework,bhattacharjee2019stratum,kourouklidis2021model,vertexwebsite,wang2020dissector}. Recently, there has been growing research interest in monitoring human-centric requirements in ML components, however, the difficulty of observing and objectively evaluating them makes it challenging. Researchers in  \cite{heyn2021requirement}, mention the importance of human factors in ML-based systems for user acceptability and trust. According to \cite{staples2016continuous}, continuous validation and monitoring for functional, non-functional and ethical user needs in ML-based systems is critical for the value of the system among users and stakeholders. Furthermore, \cite{zhu2022ai} states that the unique characteristics of ML components require continuous monitoring after deployment for human, social and environmental well-being, human-centered values, fairness, privacy protection and security. This recommendation is based on a lack of operationalised ethical ML despite the presence of several high-level frameworks \cite{zhu2022ai,mittelstadt2019principles,ibanez2022operationalising}. Based on the analysis of existing work, and commercial and open-source tools for monitoring ML components, the following human-centric requirements have been explored so far: 1) fairness; 2) trust; and 3) safety. 

Fairness refers to the property of an ML component being unbiased towards all individuals or groups. Biases in ML can be due to data disparities or inherent biases \cite{nigenda2022amazon}. There are a number of studies that monitor runtime fairness of ML components \cite{nigenda2022amazon,henzinger2023runtime,albarghouthi2019fairness}. In \cite{henzinger2023runtime}, a monitor observes the sequence of interactions with the ML component and statistically estimates the fairness of each prediction. In \cite{nigenda2022amazon}, a fairness monitor for ML models is proposed; the monitor uses statistical estimates to identify biases. A programming language for specifying ML fairness requirements and checking fairness against these requirements at runtime is presented in \cite{albarghouthi2019fairness}. Other market solutions for monitoring bias and fairness include IBM Watson's OpenScale~\cite{ibmwebsite}.

\begin{figure*}
    \centering
    \includegraphics[width=0.65\textwidth]{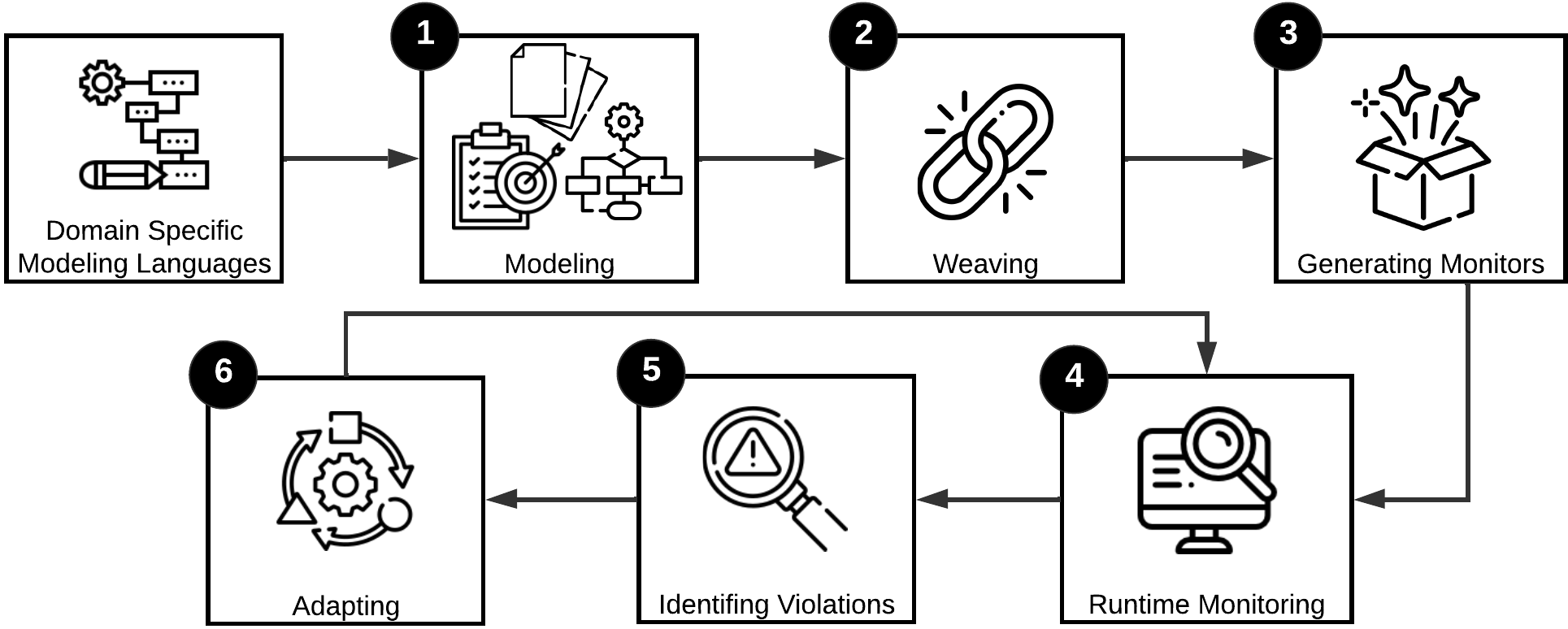}
    \caption{Proposed Approach}
    \label{fig:approach}
    \vspace*{-1em}
\end{figure*}

Trust in an ML component refers to the confidence in it to perform as expected. ML components can be monitored to assess their trustworthiness in unseen operation conditions \cite{langford2021modalas}. An approach for monitoring the trustworthiness of ML-enabled self-adaptive systems is proposed in \cite{langford2021modalas}; the study uses goal models and a deep neural network to predict the behaviour of the ML component under observation. The predictions of ML components on input data outside the training data distribution cannot be trusted \cite{byun2020manifold,roy2022runtime}. Works in \cite{byun2020manifold} and \cite{roy2022runtime} predict the probability of an input data value being an outlier and then monitor the corresponding output of the ML component. In \cite{ehrlinger2019daql}, a data quality library is presented to monitor data characteristics and maintain trust in the ML-based system.

Safety focuses on minimising risks and ensuring that the ML-based system operates within predefined safety boundaries to prevent harm to users and the environment \cite{aslansefat2020safeml}. In \cite{aslansefat2020safeml}, a monitoring approach for ML components in safety-critical systems is proposed; the study uses statistical measures to identify outliers in input data and notify the human controller. In \cite{guerin2022unifying}, new safety metrics are described for runtime monitoring of ML-based perception in autonomous systems. Monitoring the ML component in isolation is not enough to ensure safety; therefore, \cite{torfahruntime} suggests monitoring the domain boundaries or context in which the ML component operates. \cite{stocco2020misbehaviour} proposes monitoring the confidence of deep neural networks in autonomous vehicles with the help of another neural network.

Based on the state-of-the-art analysis, only a few studies monitor human-centric requirements in ML components. The ones that exist have the following key limitations: 1) Several human-centric requirements such as human values \cite{hussain2020human,nurwidyantoro2023integrating,solanki2023operationalising}, human, societal and environmental well-being \cite{haggag2022better}, and privacy \cite{haggag2021covid} were not considered. 2) Studies that propose human-centric monitoring solutions, focus only on a single human-centric requirement \cite{langford2021modalas,nigenda2022amazon,albarghouthi2019fairness}. 3) Significant effort and technical expertise in statistics and software engineering are required to set up a monitoring system~\cite{nigenda2022amazon, kourouklidis2021model}. 4) Limited support for monitoring human-centric requirements changing at runtime. For model-driven solutions for monitoring ML components, \cite{langford2021modalas} and \cite{kourouklidis2021model} are the closest to my work. However,\cite{langford2021modalas} only focuses on trust and \cite{kourouklidis2021model} does not consider any human-centric requirements. 

An important gap in the literature (that I aim to investigate in my PhD project) is the runtime monitoring of human-centric requirements in ML components \cite{zhu2022ai,ibanez2022operationalising,fellander2022achieving} to address multiple human-centric requirements, reduce technical complexities, and support automation.

\section{Proposed Solution}
I aim to address these issues in ML-based software by providing runtime monitoring of ML components for human-centric requirements violations. Early identification and correction of such issues are key for ensuring users' trust and system success. To achieve this goal, I will employ the Model-driven Engineering (MDE) paradigm and use runtime models to monitor the ML components. 
I have formulated the following key research questions for my PhD research:
\begin{itemize}
\item  \textbf{RQ1:} How can we capture end-users' human-centric requirements for systems with ML components?
\item  \textbf{RQ2:} How can we identify and map low-level technical aspects of the ML components corresponding to the human-centric requirements?
\item  \textbf{RQ3:} How can we monitor ML components at runtime for human-centric requirements violations?
\item  \textbf{RQ4:} What are the causes and context of these requirements violations?
\item  \textbf{RQ5:} Can systems with ML components self-adapt at runtime to mitigate human-centric requirements violations? If yes, how?
\end{itemize}

Fig.~\ref{fig:approach} outlines my proposed approach. A set of DSMLs are developed to specify the human-centric requirements, technical requirements, system architecture, ML system design and context. As the first step, models conforming to the DSMLs are created to capture all the aspects of the ML-based system required for runtime monitoring. Then, mappings are developed to weave the models at different levels of abstraction together. In the next step, using MDE, the models are transformed into monitors. A monitor comprises of runtime models, also known as \textit{models@runtime}, and code that facilitates continuous synchronisation between the models and the ML-based system being observed. Once the monitor is generated, synchronisation is used to monitor the behaviour of the deployed ML component in real time. An analysis is performed on the data collected from monitoring to identify any violations of the specified requirements. These violations are then classified as either fixable through self-adaptation or not fixable. The adaptation process is triggered for the former and developers are notified for the latter. 

\subsection{Modeling for Human-centric ML components} 
\subsubsection{Domain-specific Modeling} Models allow users to represent a system at a high level of abstraction while hiding technical implementation details~\cite{grundy2020humanise}. Other benefits include easier stakeholder communication, automated code generation and increased productivity\cite{grundy2020humanise,yohannis2022towards}. General purpose modeling languages such as UML \cite{eriksson2003uml} are generic and complex to apply within a particular domain \cite{grundy2020humanise}. With a DSML, users can easily capture information about their specific domain using familiar terminology \cite{france2005domain,grundy2020humanise}. In this work, a set of DSMLs will be developed to capture runtime monitoring information at different levels of abstraction. These modeling languages cover high-level human-centric requirements through low-level ML-system design and deployment. 

\textbf{Human-centric Requirements Model: } High-level human-centric requirements or goals for the ML component are specified in this model. Users can model the requirements and more detailed sub-requirements. For example, if a requirement is fairness related, the sub-requirements can be gender-based and demographics-based fairness. A sub-requirement for explainability could be visual explainability.

\textbf{Technical Requirements Model: } Low-level technical requirements for the ML component are specified in this model. Every high-level human-centric requirement would have one or more corresponding technical requirements that can be directly monitored. Similar to the human-centric requirements model, users can also model sub-requirements. For example, a technical requirement is to predict next year's sales. Another example of a technical requirement is a balanced dataset.

\textbf{System Architecture Model: } The architecture of the ML-based system is captured in this model. It includes ML components, traditional software components and the connections between them. The system architecture is required to connect the monitor to the ML component. Later, it will be used for adaptations to mitigate human-centric requirements violations.

\textbf{ML Component Design Model: } This model will include the design of the ML component and training details. For example, the ML framework used, the ML algorithm applied, evaluation metrics and hyperparameters. Design choices captured in this model will correspond to the technical requirements in the technical requirements model and an architectural component in the system architecture model. It will also support adaptations to fix human-centric requirements violations.

\textbf{Context Model: } The context in which the ML component operates has a significant impact on its behaviour \cite{heyn2021requirement}. Contextual information such as deployment details, datasets, variations between training and production data, and user interaction patterns are captured in the context model. The context model is also related to the technical requirements. Additionally, it helps developers diagnose the contextual causes for human-centric requirements violations. This is especially important in scenarios where self-adaptations are not possible.

At any point, while modeling, conflicting requirements may arise. These conflicts will be highlighted within the model so that the user can address them before moving forward. A modeling assistant can suggest the requirements that can be checked at design time and do not need runtime monitoring, however, it is beyond the scope of this PhD. 

\subsubsection{Weaving Models} 
Existing studies describe the process of mapping or linking models together as weaving\cite{del2006weaving, almorsy2014adaptable} or traceability\cite{paige2010building,galvao2007survey}. I plan to apply the same concept in my approach. Once all five models have been created in the defined DSMLs, the next step is to weave them together. It is important to map high-level human-centric requirements to low-level design aspects to facilitate monitoring. These mappings can be one-to-one, one-to-many or many-to-many. 

Consider the drone delivery system example mentioned earlier, it uses an onboard GPU-based camera and ML to identify delivery destinations. In the human-centric requirements model it has a requirement \textit{privacy of images} linked to a technical requirement \textit{recognise delivery destinations} in the technical requirements model. This, in turn, will be linked to components in the system architecture model such as an ML component \textit{destination recogniser} for computer vision tasks and a non-ML component \textit{GPU-based camera} to capture and process images. The \textit{destination recogniser} component will be linked to a design specification in the ML component design model such as a \textit{convolutional neural network (CNN)} to identify and classify delivery destinations. Finally, this \textit{privacy constrained destination recogniser} will be linked to the context such as \textit{training and testing datasets} in the context model.


\subsection{Monitor Generation} 
The next step is to apply MDE techniques \cite{khalajzadeh2020end,almorsy2014adaptable,khambati2009model,almorsy2012mdse,kim2015suite} to automatically generate monitors from the models created in the specified DSMLs. For example, a run-time privacy monitor will be generated for the destination recogniser ML component in the drone delivery system. This automation reduces the technical barriers faced by developers when setting up a monitoring system\cite{kourouklidis2021model}. As a part of my approach, two model transformers will be developed. First, a model-to-model transformer that converts the DSML models into runtime models. Second, a model-to-text transformer that generates code to continuously read data from the ML component, populate the runtime models and evaluate the runtime models against the specified requirements. A set of transformation rules will be defined and model transformers will be built to support the transformation process and enable automation. 

\subsection{Runtime Monitoring and Adaptation} 
\begin{figure}
    \centering
    \includegraphics[width=0.48\textwidth]{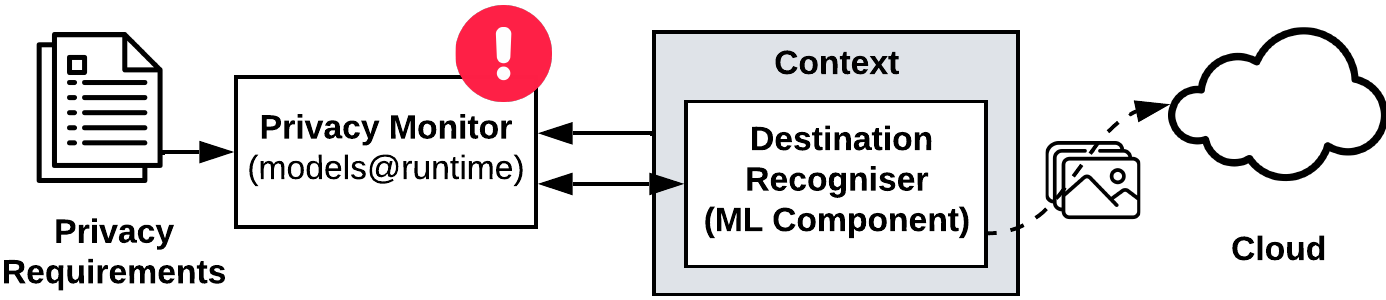}
    \caption{Privacy Monitoring of an ML Component}
    \label{fig:monitoring}
    \vspace*{-1em}
\end{figure}
During runtime, the monitor continuously collects data about the state of the ML component and its operating context. The data is stored in the runtime models and analysed against the specified requirements. The self-adaptation process will be triggered if any requirement violation is detected in the ML component. Fig.~\ref{fig:monitoring} shows privacy monitoring for the destination recogniser component of the drone delivery system. The monitor observes for any violations of privacy such as images other than the safely delivered parcel being sent to the cloud or stored in memory. If a violation occurs, the monitor identifies it and initiates self-adaptation to mitigate this violation. Some possible adaptations include obfuscating the images or shutting down the destination recogniser component. 

However, self-adaptations for all kinds of human-centric requirements violations are not possible, for example when an ML model becomes obsolete. To address this issue, I plan to classify the requirement violations into two groups. First are the requirement violations that can be fixed through self-adaptations. In this case, the adaptation process is activated when a requirement violation occurs. Second, are the violations that cannot be fixed. In this case, the developers will be notified through an alert. There are a number of existing strategies for self-adaptation, such as the MAPE-K loop \cite{arcaini2015modeling}. I plan to analyse these further to find the most suitable strategy for my work. Lastly, observability or lack of ground truth data can be a challenge for monitoring some requirements, such as fairness. For these I plan to explore metrics such as input distribution drift \cite{byun2020manifold} and prediction distribution drift \cite{ghosh2022faircanary}.

~\label{sec:solution}
\section{Evaluation and Validation Plan}
To evaluate the usefulness of the MDE approach and the ability of the monitors to detect human-centric requirements violations, the evaluation process will consist of user studies in two phases: the first will assess the prototype, and the second will evaluate the complete tool. Feedback from the prototype evaluation will be used to refine the approach and further develop the tool. To determine the effectiveness of the proposed approach, I will conduct evaluations on multiple real-world case studies. Some possible case study domains include smart homes, smart health applications, and autonomous vehicles. A few potential metrics for evaluation include reduction in technical barriers, time and effort when setting up monitors for human-centric requirements in ML components, correct and early detection of human-centric requirements violations, and correct and timely self-adaptations or alerts to developers. The primary challenge in evaluation is the subjectivity of human-centric requirements, when the monitor reports a violation, different developers may perceive it differently. To mitigate this during evaluations, I plan to provide violation evidence or explanation tailored to the users understanding \cite{zhu2022ai}. Currently, this is planned as a manual step, however, in the future, the monitors can be improved to generate explainable human-centric violation alerts for different types of users.

In case real-world case studies are not available I plan to use real-world prototypes from literature and insert human-centric requirements violations. For example, by synthesising biased input data and observing if the monitor can detect a fairness violation \cite{ghosh2022faircanary}. This mutation-based testing has been successfully applied to evaluate ML-based systems \cite{ma2018deepmutation,humbatova2021deepcrime}.

~\label{sec:evaluation}

\section{Current Status}

Table 1 summarises my PhD timeline. I have completed data analysis for a systematic literature review on MDE for systems with ML components. While conducting the literature search, I found only a few studies considering human-centric aspects while providing MDE solutions for systems with ML components. Fewer studies are available that leverage MDE for monitoring systems with ML components. To this end, I have reviewed existing literature on monitoring approaches for ML and identified the need for an MDE approach to monitor human-centric requirements in ML components. 

\begin{table}[htb] 
\vspace*{-1em}
\centering
\caption{PhD timeline }
\vspace*{-0.5em}

\footnotesize
\begin{tabular}{ p{5.5cm}  p{2.5cm}} 
\hline
 \TBstrut \textbf{Research Task} & \textbf{Duration} \\ [0.5 ex]
\hline
 \TBstrut Systematic literature review and problem identification & Feb 2023 - Jul 2023\\
\hline
 \TBstrut Identifying and analysing human-centric requirements &  Aug 2023 - Jan 2024 \\
\hline
 \TBstrut Analysis and selection of adaptation mechanism & Feb 2024 - Apr 2024 \\
\hline
 \TBstrut Solution formulation, refinement and prototype implementation & May 2024 - Oct 2024 \\
\hline
 \TBstrut Prototype evaluation &  Nov 2024 - Feb 2025 \\
\hline
 \TBstrut Further solution refinement and tool development & Mar 2025 - Jul 2025 \\
\hline
 \TBstrut Evaluation on real-world case studies & Aug 2025 - Jan 2026 \\
\hline
 \TBstrut Thesis writing and defence & Feb 2026 - Jul 2026 \\
\hline 
\\
\end{tabular}
\vspace*{-1em}
\end{table}

I next intend to do surveys with users and practitioners of ML-based systems to identify the key human-centric requirements that necessitate runtime monitoring. After identifying the requirements, I will analyze their monitoring feasibility and select the ones that can be addressed within the scope of this PhD. Subsequently, I will compare existing adaptation mechanisms to determine the most suitable one for my research. Once this is done, I will proceed to the detailed development of a solution based on the information gathered in the previous steps. Next, I will implement and evaluate the prototype, refine it, and build a tool for evaluations on real-world case studies. 

~\label{sec:status}

\section*{Acknowledgements}

Naveed is supported by a Faculty of IT PhD scholarship. This work is partially supported by ARC Discovery Project DP200100020 and ARC Laureate Fellowship FL190100035.

\bibliographystyle{IEEEtran}

\bibliography{references}

\end{document}